\documentclass[aip,twocolumn,superscriptaddress]{revtex4}
\usepackage{amsmath,amssymb,color}
\usepackage{graphicx}
\usepackage{dcolumn}
\usepackage{bbm}
\usepackage{bm}
\usepackage{pifont}

\begin{document}

\title[THE JOURNAL OF CHEMICAL PHYSICS, XXX, XXXXXX (201X)]{Normal versus anomalous self-diffusion in two-dimensional fluids: Memory function approach and generalized asymptotic Einstein relation}

\author{Hyun Kyung Shin}  
 \affiliation{ 
Department of Chemistry, Korea Advanced Institute of Science and Technology, Daejeon 305-701, Republic of Korea}

\author{Bongsik Choi}  
 \affiliation{ 
Department of Chemistry, Korea Advanced Institute of Science and Technology, Daejeon 305-701, Republic of Korea}

\author{Peter Talkner}
 \affiliation{Universit{\"a}t Augsburg, Institut f{\"u}r Physik, D-86135 Augsburg, Germany}%
\affiliation{Asia Pacific Center for Theoretical Physics (APCTP), San 31, Hyoja-dong, Nam-gu, Pohang, Gyeongbuk 790-784, Korea}
 
\author{Eok Kyun Lee}%
 \thanks{Corresponding author}
 \email{eklee@cola.kaist.ac.kr} 
 \affiliation{ 
Department of Chemistry, Korea Advanced Institute of Science and Technology, Daejeon 305-701, Republic of Korea}

\date{\today}

\begin{abstract}
Based on the  generalized Langevin equation for the momentum of a Brownian particle a generalized asymptotic Einstein relation is derived. It agrees with the well-known Einstein relation in the case of normal diffusion but continues to hold for sub- and super-diffusive spreading of the Brownian particle's mean square displacement. The generalized asymptotic Einstein  relation is used  to analyze data obtained from molecular dynamics simulations of a two-dimensional soft disk fluid. We mainly concentrated on medium densities for which we found super-diffusive behavior of a tagged fluid particle. At higher densities a range of normal diffusion can be identified. The motion presumably changes to sub-diffusion for even higher densities.      

\end{abstract}

\maketitle

\section{Introduction}

Brownian motion, as a theory for the dynamics of a particle immersed in a resting fluid, has played a substantial role in the development of the statistical mechanics of diverse fluctuation phenomena both in and out of equilibrium situations \cite{Zwanzig,HM}. Even though a Brownian particle rests on average, it explores with time ever larger parts of space  if not hindered by confining walls or other spatial constrifctions. This spreading can be quantified in terms of the particle's mean square displacement $\langle (\delta X(t))^2 \rangle$  which, in a resting, three-dimensional fluid at thermal equilibrium, grows at large times $t$ proportionally to $t$. In this case of so-called {\it normal} diffusion, the diffusion constant $D$ defined as the proportionality factor in the spreading law $\langle (\delta X(t))^2 \rangle = D t $ is related to the friction coefficient $\gamma$ experienced by the Brownian particle in the fluid and the temperature of the fluid by the Einstein relation \cite{Einstein}
\begin{equation}
D=\frac{2 k_B T}{M\gamma}
\label{ERnd}
\end{equation}
This relation actually goes back to Sutherland \cite{Sutherland}. 

Other forms of diffusion laws known as {\it anomalous} diffusion may occur in the presence of spatial constriction, or in non-equilibrium situations \cite{BG,MK}. Often, the spreading may still be described by a power-law in time with an exponent differing from one, $\langle (\delta X(t))^2 \rangle = D_\alpha t^\alpha$. For example, in so-called single file diffusion, the location of a hard sphere in a one-dimensional row of other equal impenetrable spheres  spreads with the exponent $\alpha=1/2$ \cite{AP}. In extreme contrast to this dwindling dispersion of single file diffusion, the distance $R$ of a pair of particles which are advected in a turbulent fluid in three dimensions spreads according to Richardson's law, $\langle R^2(t) \rangle \propto t^3 $, \cite{R,B}, with the exponent $\alpha=3$.
  
In a two-dimensional fluid at equilibrium, mode-coupling theory predicts the diffusive motion of a Brownian particle as well as the self-diffusion of fluid particles to follow a logarithmic correction to the algebraic behavior of the form $\langle (\delta X(t))^2 \rangle \propto t \ln t$ \cite{AW}. A selfconsistent mode-coupling theory yields a slightly weaker increase  given by $\langle (\delta X(t))^2 \rangle \propto t \sqrt{\ln} t$ \cite{K,WAG}. 

For the motion of a Brownian particle, and also of a tagged fluid particle a formally exact description exists in terms of a generalized Langevin equation, which constitutes a linear integro-differential equation for the considered particle's momentum. Given the molecular interactions, the memory kernel and the fluctuating force which are the ingredients of the generalized Langevin equation, can formally be derived by using the projection operator techniques of Zwanzig \cite{Zwanzig2} and Mori \cite{Mori}. The nature of the resulting motion of the Brownian particle is determined by the behavior of the integral of the memory kernel.  
The motion is diffusive at large times if the time-integral of the memory kernel converges in the upper limit to a finite, non-vanishing value, called the static friction. It becomes super-diffusive, i.e., roughly speaking, the mean square displacement grows with  an exponent $\alpha >1$, if this integral vanishes, and the motion is sub-diffusive ($\alpha <1$) if the static friction diverges.
These properties follow from the mutual dependence of the memory kernel and the momentum autocorrelation function at large times as first found by Corngold \cite{Corngold}.

Here we demonstrate that, at large times, properly defined time-dependent friction and diffusion coefficients stay in a reciprocal relation to each other. Their product is determined by a generalized asymptotic Einstein relation which, for normal diffusion, agrees with its well-known form (\ref{ERnd}).  Most notably, the generalized relation allows one to determine the scaling exponent $\alpha$ of the particle's mean square displacement. This is the main difference between the present, generalized asymptotic Einstein relation and previous forms of generalized Einstein relations \cite{BG,BF} which are not restricted to the asymptotic large-time regime but do not explicitly contain the scaling exponent.

Unfortunately, the exact molecular expressions for the memory kernel and the fluctuating force are extremely involved and most often cannot be analytically determined other than in limiting cases. Molecular dynamics (MD) simulations     
provide an alternative, convenient means to study the Brownian motion as well as the motion of a tagged fluid particle. As an example, we investigate a fluid in two spatial dimensions modeled by $N$ soft spheres interacting pairwise via purely repulsive Lennard-Jones potentials.   

The paper is organized as follows. In Section II, the microscopic model
and the parameters for the molecular dynamics simulation are specified. After a short review of the generalized Langevin equation, the generalized asymptotic Einstein relation is derived from the generalized Langevin equation by means of the Tauberian theorem in Section III. For the sake of completeness and also for comparison, we present the proof of the already mentioned  generalization of the Einstein relation \cite{BG,BF} at the end of the same Section. In Section IV the scaling behavior of the time-dependent diffusion and friction coefficients are determined in the framework of the molecular dynamics simulations. The scaling exponents are found to be compatible with each other in the sense that they concordantly imply the same scaling exponent $\alpha$ for the mean square displacement. We also find good agreement with the value of $\alpha$ following from the generalized asymptotic Einstein relation. The slight difference most likely can be attributed to the presence of slowly varying functions which may modify the apparent algebraic scaling behavior. Because from the available data the scaling behavior of the diffusion and friction coefficients is visible only for relatively short times it is impossible to identify and separate the  contribution of such a possibly existing, slowly varying function to the apparent scaling. In the generalized asymptotic Einstein relation the slowly varying contributions of the diffusion and friction coefficients compensate each other. This effect leads to a more reliable estimate of the exponent $\alpha$. The paper closes with concluding remarks and a discussion of the present findings to previously published results in Section V.

\section{Microscopic model and Molecular dynamics simulation method}\label{MD}

We considered a standard model of a Brownian particle suspended in a two-dimensional fluid \cite{AT}. It consists of a single probe particle of mass $M$ and diameter ${\sigma}_\textrm{BB}$ and $N$ solvent particles of mass $m$ and diameter ${\sigma}_{\textrm{SS}}$ enclosed in a two-dimensional domain of side-length $L_{x}$ and $L_{y}$ with periodic boundary conditions. 

Assuming that the system is isolated, the motion of particles follows the Hamiltonian dynamics with the Hamiltonian given as
\begin{equation}\label{Hamiltonian} 
{\mathcal H} = \sum_i \frac{1}{2m}\left|{\mathbf p}_i \right|^2 + \frac{1}{2M}\left|{\mathbf P}\right|^2 + U \left({\mathbf r}^N, {\mathbf R} \right)\:,
\end{equation}
where ${\mathbf r}_i$ and ${\mathbf p}_i$ are the position and the momentum of the $i$th solvent particle. Accordingly, ${\mathbf R}$ and ${\mathbf P}$ refer
to the probe particle, and $U({\mathbf r}^N,{\mathbf R})$ is the  potential energy describing the mutual interaction of the solvent particles among each other and with the Brownian particle. 
The total linear momentum of the system as well as its energy are conserved. The angular momentum is not conserved due to periodic boundary condition. This assumption applies to the standard MD model and generates the NVEp ensemble for which the number of particles, volume, total energy and total linear momentum are fixed.

The potential energy is determined by pairwise interactions and hence given by
\begin{equation}
U(\mathbf{r}^N,\mathbf{R}) = \sum_{i>j} u_{SS}(|\mathbf{r}_i -\mathbf{r}_j|) + \sum_i u_{BS}(|\mathbf{R}-\mathbf{r}_i|)\;.
\label{Uuu}
\end{equation}
For the pair-potentials $u_{SS}(r)$, and $u_{BS}(r)$, we used  
purely repulsive Lennard-Jones potentials \cite{Weeks} of the form 
\begin{equation}\label{WCA_ER2D}
u_{\alpha \beta}(r)=\left\{ \begin{array}{ll}
4\epsilon \left[\left({\sigma}_{\alpha \beta}/r\right)^{12} - \left({\sigma}_{\alpha \beta}/r\right)^{6} \right] + \epsilon  & \mbox{for $r < 2^{1/6} {\sigma}_{\alpha \beta}$} \\ 
0 & \mbox{for $r \geq 2^{1/6} {\sigma}_{\alpha \beta}$}. \end{array}
\right.
\end{equation}
where $\alpha$ and $\beta$ stands for either $S$ or $B$ referring to the fluid and Brownian particles, respectively. The length-scale ${\sigma}_\textrm{SB}=\left({\sigma}_{\textrm{BB}}+{\sigma}_{\textrm{SS}}\right)/2$  determines the range of the repulsive fluid-Brownian-particle interaction.

In the MD simulations, dimensionless units were used. The lengths are measured in units of $\sigma_{SB} $. 
For the number density $n$ of fluid particles we took $n \sigma_{SB}^2 =0.3,\; 0.4,\;0.5,\;0.8$. 
With the choice of the fluid particle size,  the diameter $\sigma_{BB}$ of the Brownian particle is also determined. 
The total number $N$ of fluid particles was taken as $N = 32\:000$ in the majority of cases. In order to identify finite size effects, simulations with less particles were performed at constant number density by taking $N=160,\; 1\:000,\;4\:000,\;16\:000$. The configuration space consisted of a rectangle glued together at opposite sides. The side lengths were  chosen as $L_x= \sqrt{n/N} (N/n + \pi \sigma^2_{BB}/4)$ and $L_y=\sqrt{N/n}$. Note that, for the considered numbers of particles, the deviation from a square is only minor.

All masses were measured in units of the fluid particle mass $m$, and energies in units of the strength $\epsilon$ of the pair interactions at the distance $r=\sigma_{\alpha \beta}$. 
A consistent unit of time then follows as $\tau \equiv \sigma_{SB} \sqrt{m/\epsilon}$.

As initial conditions of the MD simulations the centers of the fluid particles were put on the points of a simple cubic (100) surface and the Brownian particle was positioned at the center of one of the squares formed by four neighboring fluid particles. For each solvent particle  a vector $\sqrt{k_B T^{\text{set}}/m}\: \mathbf{e}$  was chosen, where $k_B$ denotes the Boltzmann constant and $T^{\text{set}}$ the target temperature. The two-dimensional vector $\mathbf{e}$ was taken randomly  with independent, Gaussian distributed Cartesian components having vanishing averages and unit variances. The algebraic average over all these $N$ vectors was subtracted from the individual vectors. The resulting vectors were used as initial velocities of the fluid particles. The initial velocity of the Brownian particle was put to zero.
By construction, the total momentum of the  system consisting of fluid and Brownian particles vanished.  
The values of target temperature $T^{\text{set}}$ were taken as $0.2 \epsilon / k_B$, $0.5 \epsilon / k_B$  
and  $ 1.0 \epsilon / k_B$.

The equations of motion for each individual particle were solved using the velocity Verlet algorithm with a time step $\Delta t = 0.001\tau $. 
At the beginning of the simulation, a velocity rescaling was performed every $10,000$ steps to adjust the energy of the system to the desired value (all together $10$ times). 
We found that the thermal equilibrium was established within $2\times 10^{6}$ time steps. 
We used the data resulting from the subsequent $1 \times 10^{7}$ time steps for the construction of the time correlation functions of the relevant dynamical variables and the GLE.
If necessary,  $20$ simulations were performed for identical thermodynamic parameters to obtain ensemble-averaged quantities. 

\section{Generalized Mori-Zwanzig Langevin equation for a probe particle in a 2D fluid}\label{GLE}
\subsection{Generalized Langevin equation}
According to the Mori-Zwanzig theory \cite{Zwanzig} the dynamics of the probe particle can be represented as a generalized Langevin equation for the momentum ${\mathbf P}(t)$ reading 
\begin{equation}\label{GLE_tensor} 
\frac{d}{dt}{\mathbf P}(t) = {\mathbf \Omega}\cdot {\mathbf P}(t)-\int^t_0{{\mathbf k}(t^\prime)\cdot {\mathbf P}\left(t-t^\prime\right) dt^\prime }+{\mathbf F}^\dagger(t)\:. 
\end{equation}

The first term on the right hand side describes the reversible, instantaneous change of the two-dimensional momentum $\mathbf{P}(t)$ with Cartesian components $P_x(t)$ and $P_y(t)$. It is determined by the frequency matrix
$ \mathbf{\Omega}$, which is defined as
\begin{equation}\label{omega_GLE}
{\mathbf \Omega } \equiv \left({\mathbf P}, L{\mathbf P}\right)\cdot {\left({\mathbf P},{\mathbf P}\right)}^{-1}\:,
\end{equation}
where $\left (f,g \right ) =$ \\
$ \int d \Gamma f(\Gamma) g(\Gamma) e^{- \mathcal{H}(\Gamma)/( k T^{\text{set}})} /\int d \Gamma e^{- \mathcal{H}(\Gamma)/( k T^{\text{set}})}$ denotes the Mori scalar product of phase-space functions $f(\Gamma)$ and $g(\Gamma)$ with $\Gamma$ being a point in phase-space  and $d\Gamma= d^N \mathbf{p}d^N \mathbf{r} d \mathbf{P} d \mathbf{R}$ the corresponding infinitesimal phase-space volume element. This scalar product yields the equilibrium correlation of its arguments.
The Liouville operator is defined as $ L f= \{H,f \} $, where $\{f,g\}$ denotes the Poisson bracket of the phase-space functions $f$ and $g$. Because the weight used in the Mori scalar product is stationary, the frequency matrix $\Omega$ vanishes. 
A projection operator $\mathcal{P}$ is conveniently defined in terms of the Mori scalar product assigning to each phase-space function $f(\Gamma)$ the component parallel to $\mathbf{P}$ as
\begin{equation}
\mathcal{P} f = \mathbf{P} \cdot \left (\mathbf{P}, f \right )\:,
\label{PP}
\end{equation}   
where the dot indicates the scalar product of two-dimensional vectors $\mathbf{U} \cdot \mathbf{V} = U_x V_x +U_y V_y$.  
The fluctuating force $\mathbf{F}^\dagger(t)$ can then be expressed as
\begin{equation}\label{fluc_GLE} 
{\mathbf F}^\dagger(t) \equiv \left({\mathbf 1}-{\mathcal P}\right) \exp{\left\lbrace \left({\mathbf 1}-{\mathcal P}\right) Lt \right\rbrace} L{\mathbf P}\:.
\end{equation}
Its mean value vanishes, $\langle \mathbf{F}^\dagger(t) \rangle =0$, where the brackets indicate a thermal equilibrium average.
The memory kernel $\mathbf{k}(t)$ is defined in terms of the fluctuating force-force correlation function as
\begin{equation}\label{kernel_GLE}
{\mathbf k}(t) \equiv \left({\mathbf F}^\dagger,{\mathbf F}^\dagger(t)\right)\cdot {\left({\mathbf P},{\mathbf P}\right)}^{-1}\:. 
\end{equation}
Due to the isotropy of the present model \cite{aniso}  , the memory kernel is proportional to the $2 \times 2$ unit matrix $\mathbbm{1}$, i.e. it becomes
\begin{equation}
\mathbf{k}(t) = k(t) \mathbbm{1}\:.
\label{kiso}
\end{equation}
Hence, the Cartesian components of the momentum do not couple to each other and obey the identical generalized Langevin equations of the form
\begin{equation}
\frac{d}{dt} P_\alpha(t) = -\int_0^t k(t') P_\alpha(t-t') dt' +F^\dagger_\alpha(t)\:,\quad \alpha=x,y\:.
\label{GLE_alpha}   
\end{equation}
Performing  the Mori scalar product with the momentum component $P_\alpha$ on both sides of this equation one obtains for the momentum autocorrelation function (MACF) $C(t) \equiv (P_x,P_x(t) ) =(P_y,P_y(t) ) $
a homogeneous integro-differential equation reading \cite{Zwanzig}

\begin{equation}\label{GLE_MACF}
\frac{d}{dt}C \left(t\right) = -\int^{t}_0{k \left(t^\prime\right) C \left(t-t^\prime \right)dt^\prime}\:.  
\end{equation}
The inhomogeneity vanishes because of the orthogonality of the initial momentum and the fluctuating force, i.e., $( P_x, F^\dagger_x(t)) = ( P_y, F^\dagger_y(t)) =0$.

From the knowledge of the instantaneous 
momenta of 
the probe particle, we estimated the stationary momentum auto-correlation function $C(t)$. In principle, knowing  $C(t)$ one can find the memory kernel by solving Eq.~(\ref{GLE_MACF}) for $k(t)$. 
A straightforward formal solution is obtained by means of a Laplace transformation which yields
\begin{equation}
\hat{C}(z) = \frac{C(0)}{z+\hat{k}(z)}\:,
\label{Cz}
\end{equation}
where $\hat{C}(z)= \int_0^\infty e^{- z t} C(t)$ and $\hat{k}(z)= \int_0^\infty e^{- z t} k(t)$.
However, the inverse Laplace transform poses notorious numerical problems and therefore one has to resort to Eq.~(\ref{GLE_MACF}) in the time regime. 
This amounts to the solution of a homogeneous Volterra equation of the first kind, which is also known for its numerical instability. We therefore pursued a more stable approach as described in Ref. \cite{Shin}. For this purpose Eq.~(\ref{GLE_MACF}) is differentiated with respect to time yielding
\begin{equation}\label{GLE_MACF_diff}
\ddot{C}(t) = -C(0)k(t)-\int^t_0{k(t^\prime)\dot{C} \left(t-t^\prime\right)dt^\prime}\:,
\end{equation} 
which is a Volterra equation of the second kind. 
The first derivative of the MACF is given by the stationary cross-correlation between the momentum and its first derivative with respect to time. Therefore it agrees with the cross-correlation of the momentum and the total force $F_\alpha(t) \equiv -\int_0^t k(t') P_\alpha(t-t') + F^\dagger_\alpha(t)$ acting on the Brownian particle yielding
\begin{equation}\label{MACF_diff_GLE}
\dot{C}(t) = \left(P_x, F_x(t) \right) =\left(P_y, F_y(t) \right)\:.
\end{equation}
Likewise, the 
second derivative of the MACF is given by the negative autocorrelation function of the total force acting on the Brownian particle, such that
\begin{equation}\label{MACF_diff_diff_GLE}
\ddot{C}(t) = -\left( F_x, F_x(t) \right) = -\left( F_x, F_x(t) \right)\:. 
\end{equation}
Because the total force exerted on the Brownian particle can be determined from the MD simulation, also the first and the second derivative of the MACF can be estimated from the data in terms of the correlation functions $(P_\alpha,F_\alpha(t))$ and $(F_\alpha,F_\alpha(t))$ without invoking numerical differentiation. Finally, the memory kernel can reliably be determined from Eq.~(\ref{GLE_MACF_diff}) in a numerically stable way \cite{Shin}.

\begin{figure}\begin{center}
\includegraphics[width=8.5cm]{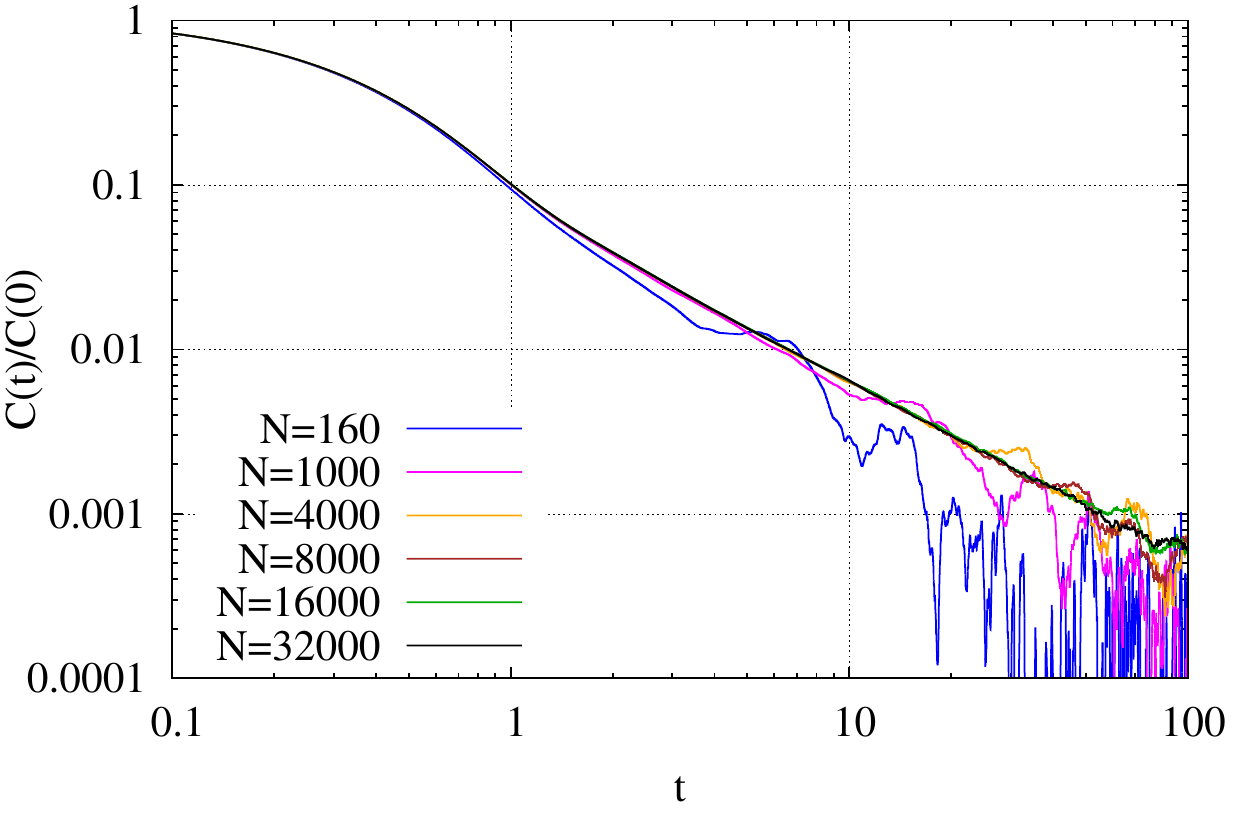}
\caption{The MACF $C(t)$ of a probe particle of the same size and mass as the fluid particle is displayed as a function of time in a doubly logarithmic presentation for different system sizes, $N=160$ (brown), $N=1,000$ (magenta), $N=4,000$ (orange), $N=8,000$ (green), $N=16,000$ (blue), and $N=32,000$ (red) are plotted. Density and temperature of the fluid are $n=0.4$ and $T^{\text{set}}=1.0$, respectively. The approach to an algebraic behavior at large times becomes more pronounced with increasing system size.}
\label{VACF_N}
\end{center}\end{figure}

\begin{figure}\begin{center}
\includegraphics[width=8.5cm]{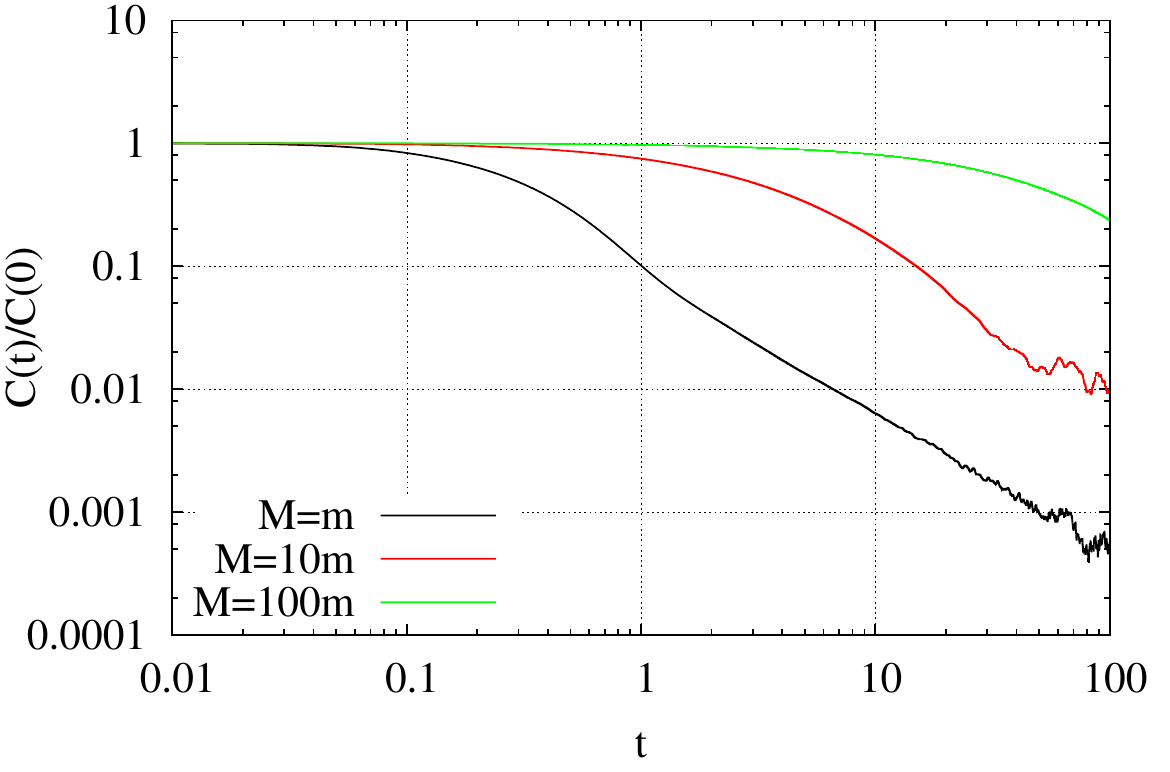}
\caption{The MACF $C(t)$ of a probe particle of the same size  as the other $N=16,000$ fluid particles is displayed as a function of time $t$ in a doubly logarithmic plot. The density and temperature are chosen as $n=0.4$ and $T^{\text{set}}=1.0$, respectively. Three cases with different solute masses, $M=m$ (black), $M=10 m$ (red), and $M=100m$ (green),  are compared. 
With increasing mass the MACF decays slower with a decay law that gradually changes from an algebraic  to a more exponential-like behavior at large times.}
\label{VACF_loglog_M___}
\end{center}\end{figure}

\begin{figure}\begin{center}
\includegraphics[width=8.5cm]{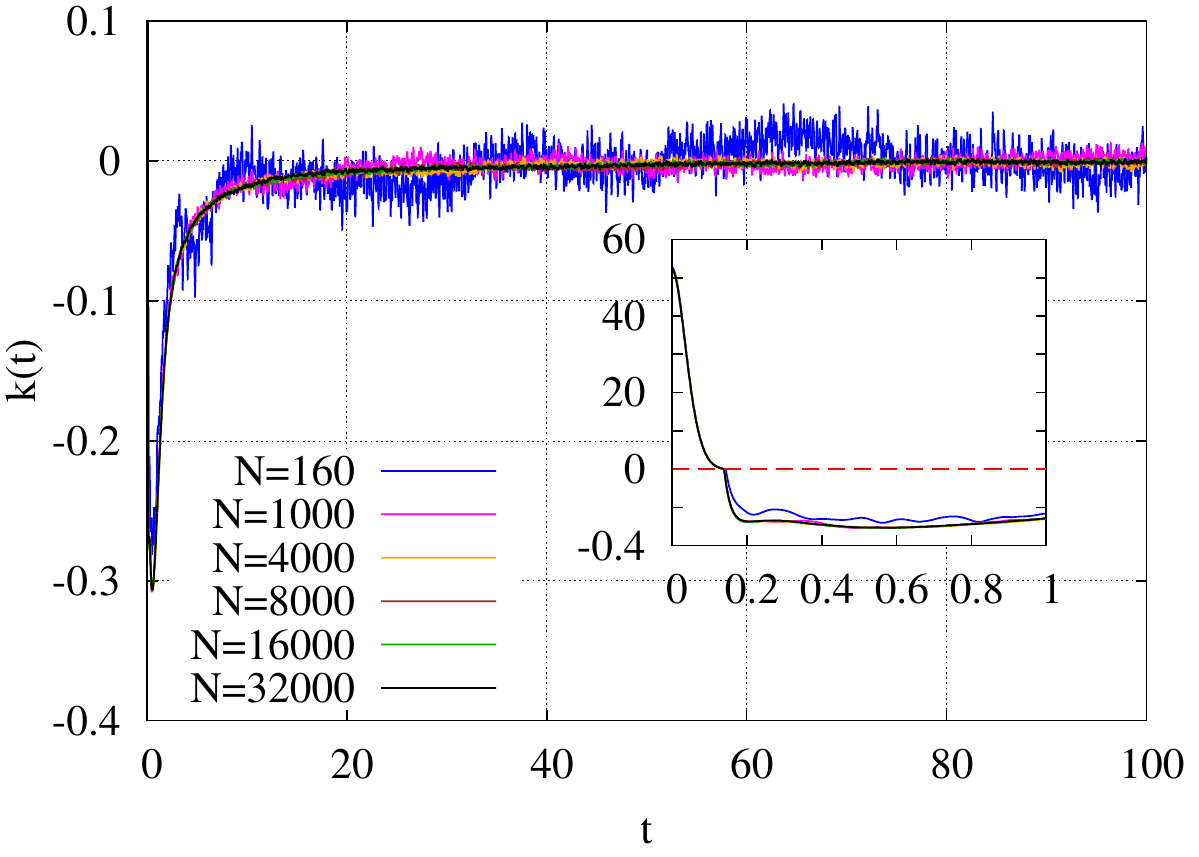}
\caption{The memory function $k(t)$ is presented as a function of time for  self-diffusion in a fluid at the density $n=0.4$ and temperature $T^{\text{set}}=0.6$ for a fluid consisting of  $N=160$ (blue), $N=1\:000$ (red), $N=4\:000$ (yellow), $N=8000$ (magenta), $N=16\:000$ (green) and $N=32\:000$ (black) particles. As for the MACF the memory functions become less rugged the larger the system size becomes. Differences between the memory functions for $N=16\:000$ and $N=32\:000$ are hardly visible. The inset displays the short-time behavior of the memory function. For better visibility, the negative part is magnified by a factor of 50. A dependence of the memory kernel on the system size $N$ becomes only sensible for larger times when $k(t)$ becomes negative.}
\label{F_ktN}
\end{center}\end{figure}

\begin{figure}\begin{center}
\includegraphics[width=8.5cm]{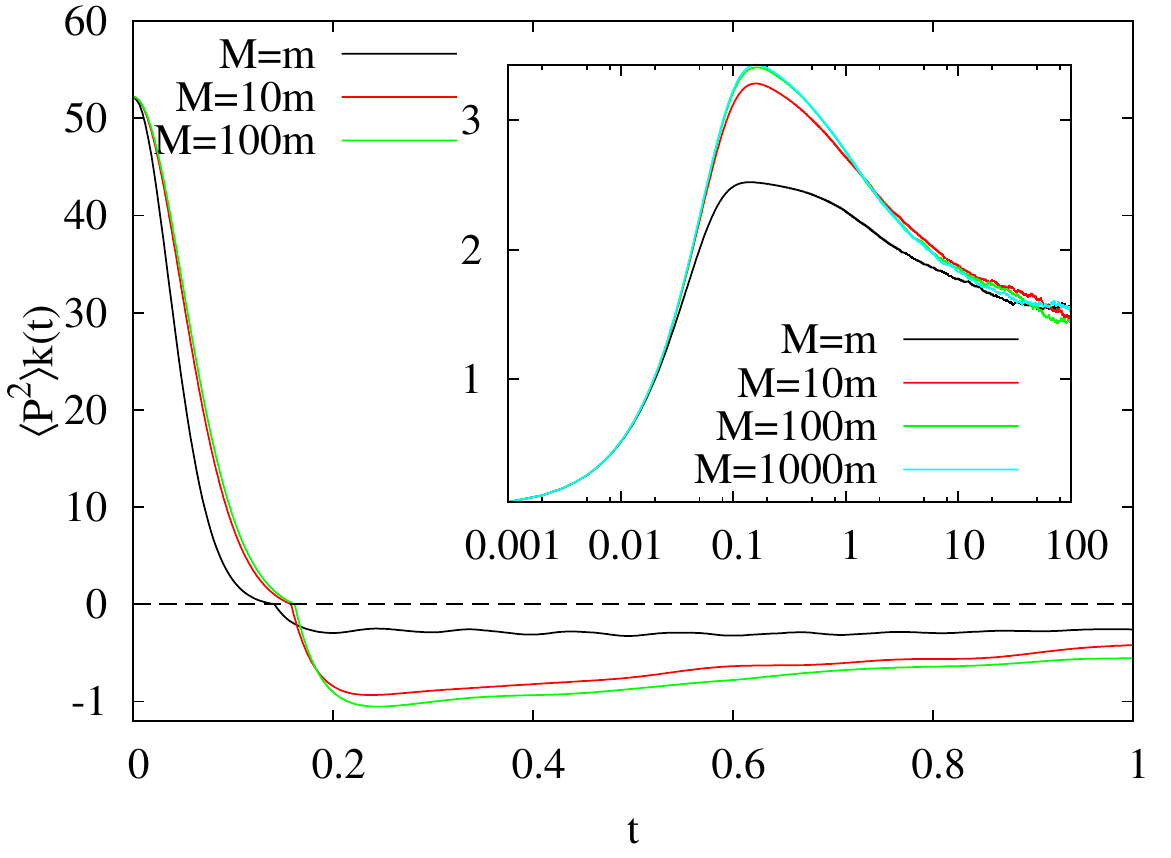}
\caption{The memory function $\langle P^2\rangle k(t)$ of the solute particle is presented for short times. 
As in Fig.~\ref{VACF_loglog_M___}, three different masses with $M=m$ (black), $M=10m$ (red), and $M=100m$ (green) are displayed.
For the sake of better visibility, in the region where the memory function becomes negative its value is magnified by a factor of 10.
With increasing mass the decay gradually becomes slower  in the region where the memory function is positive and, at the same time, the negative part becomes more pronounced.
This trend corresponds to the result displayed in the inset which displays the  integrated memory kernel as a function of time. The larger the mass the more pronounced is the maximum of memory kernel.}
\label{kernel_M___}
\end{center}\end{figure}

Fig.~\ref{VACF_N} displays the dependence of the MACF on the system size for self-diffusion ($M=m$) in a fluid with density $n=0.4$ and $T^\textrm{set}=1.0$.
Here and in the following we use the dimensionless units as introduced in Section II. At large times for the largest system size $N=32\:000$ the MACF follows an algebraic decay approximately as $t^{-1}$ with some small bumpy deviations at the largest displayed times. These deviations increase in size and move to shorter times with decreasing $N$. 
Fig.~\ref{VACF_loglog_M___} and Fig.~\ref{kernel_M___} exemplify  the MACFs and corresponding memory functions of the Brownian particle at the same density and temperature as in Fig.~\ref{VACF_N}, but with different solute masses. 

The memory functions  were numerically determined from Eq.~(\ref{GLE_MACF_diff}) by means of the estimated momentum-total-force and total-force-total-force correlation functions (\ref{MACF_diff_GLE}) and (\ref{MACF_diff_diff_GLE}), respectively. The obtained memory functions are characterized by a rapidly decaying peak at short times, then cross the zero line and only slowly re-approach zero from the negative side.
Fig.~\ref{F_ktN} exemplifies this behavior for self-diffusion ($M=m$, $\sigma_{BB}=\sigma_{SS}$) at the density $n=0.4$, and temperature $T^{\text{set}}=1.0$ for different system sizes. It is interesting to note that the positive short-time part of the memory kernel is virtually unaffected by the system size. This positive part can be understood as resulting from  sequences of independent binary collisions between the tagged particle and the fluid particles. Hence it does not contain finite size effects. At larger times when the memory kernel becomes negative correlations between collisions may build up in a way depending on the system size. 
The larger the system size the smoother the memory kernel becomes. 

Fig.~\ref{kernel_M___} demonstrates the behavior of the memory kernel for different masses at the same density and temperature for $N=16\:000$. The inset of this figure displays the integrated memory kernels, or friction kernels as we will denote them below. They continue to decrease until the largest times without reaching a plateau value possibly indicating sub-diffusive behavior.

\subsection{Generalized Einstein Relation}
The position of a Brownian particle in an isotropic medium typically spreads at large times according to an algebraic law imposing for the variance $\langle (\delta X(t))^2 \rangle$ a growing as 
\begin{equation}
\langle (\delta X(t))^2 \rangle \sim D_\alpha t^\alpha \quad \text{for}\; t \to \infty\:, 
\label{Dalpha}
\end{equation}
where $\delta X(t) = X(t) -X(0)$ is the fluctuation of a component of the Brownian particle position $\mathbf{R}(t)$, and $\alpha>0$ is a scaling exponent. Possibly existing ``slow'' corrections to the algebraic law will be considered later. In the simplest case described by Eq.~(\ref{Dalpha}) the diffusion is {\it normal} for $\alpha =1$, {\it sub-diffusive} for $\alpha < 1$ and {\it super-diffusive} for $\alpha>1$. By the $\sim$ symbol we indicate the asymptotic equality of two functions $f(t)$ and $g(t)$, i.e. $f(t) \sim g(t)$ for $t \to T$ if $\lim_{t \to T} f(t)/g(t) =1$. 

Because of $\delta X(t) = M^{-1} \int_0^t ds P(s)$, where $P(t)$ is the same component of the momentum vector $\mathbf{P}(t)$ as $X(t)$ of the position vector $\mathbf{R}(t)$, the variance of the position is related to the momentum autocorrelation  by $\langle (\delta{X}(t))^2 \rangle = M^{-2} \int_0^t ds_1 \int_0^t ds_2  C(s_1-s_2)$. Introducing the time-dependent {\it diffusion coefficient} $D(t)$ as the time-derivative of the position variance we obtain
\begin{equation}
\begin{split}
D(t) & \equiv \frac{d}{dt} \langle (\delta X(t))^2 \rangle \\
& = \frac{2}{M^2} \int_0^t ds\: C(s)\:.
\end{split}
\label{Dt}
\end{equation}  
The Laplace transform of the diffusion coefficient, $\hat{D}(z) = \int_0^\infty D(t) e^{-zt}$, is consequently related to the Laplace transform of the momentum autocorrelation by
\begin{equation}
\hat{D}(z) = \frac{2}{M^2 z} \hat{C}(z)\:.
\label{Dz}
\end{equation}
Introducing the time-dependent {\it friction coefficient} 
\begin{equation}
\gamma(t) \equiv \int_0^t ds \: k(s)\:,
\label{gt}
\end{equation}
and, accordingly, its Laplace transform
\begin{equation}
\begin{split}
\hat{\gamma}(z) &\equiv \int_0^\infty \gamma(t) e^{-z t}\\
&= \frac{1}{z} \hat{k}(z)\:,
\end{split}
\label{gz}
\end{equation}
we obtain from Eq.~(\ref{Cz})
\begin{equation}
z^2 \hat{D}(z)(1+\hat{\gamma}(z)) = 2 C(0)/M^2\:.
\label{zDg}
\end{equation}

In order to elucidate the physical meaning of the time-dependent friction coefficient we consider a weak external, time-dependent force $F_e(t)$ acting on the Brownian particle. Within the linear response regime, the dynamics of the particle is then governed by the generalized Langevin equation
\begin{equation}
\dot{P}(t) = -\int_0^t ds k(t-s) P(s) +F_e(t) + F^+(t)\:.
\label{PFe}
\end{equation}
For the sake of simplicity, the direction of the force is supposed to be constant in time and only the motion in the direction of the external force is considered. Averaging over the realizations of the fluctuating force we obtain a linear equation for the mean value of the momentum governed by the memory kernel and the applied external force reading
\begin{equation}
\langle \dot{P}(t) \rangle_e = - \int_0^t ds k(t-s) \langle P(s) \rangle_e + F_e(t)\:,
\label{mPFe}
\end{equation}
where $\langle \cdot \rangle_e$ denotes an non-equilibrium average in the presence of the driving force $F_e(t)$. 
From the mean value equation (\ref{mPFe}) it follows that the Brownian particle moves with  constant mean momentum through the fluid, provided that the external force is proportional to the integral of the memory function, i.e. proportional to the time-dependent friction coefficient: 
\begin{equation}
F_e(t) \propto \gamma(t) \quad \Longrightarrow \langle P(t) \rangle_e = \text{constant}\:.
\label{Fg}
\end{equation}
This generalizes the Archimedean law of uniform motion at constant force in a medium that causes normal diffusion  to achieving a uniform motion of a particle in an environment that causes anomalous diffusion.
  
    
To further elucidate the relation between the diffusion and friction coefficients at large times  we will investigate  Eq.~(\ref{zDg}) for small, positive values of the Laplace variable $z$. 
For the asymptotic behavior of the time-dependent diffusion coefficient we consider the slightly more general form 
\begin{equation}
D(t) \sim t^{\alpha -1} L(t)
\label{DaL}
\end{equation}
than the one that immediately follows from the purely algebraic dependence by differentiating   Eq.~(\ref{Dt}) with respect to time. This relation agrees in the scaling exponent $\alpha -1$ with the time-derivative of (\ref{Dt}) but, additionally, contains a slowly varying function $L(t)$, i.e. a function satisfying the asymptotic relation $\lim_{t\to \infty} L(t x)/L(t) =1$ for all positive $x$ \cite{F}. That means that $L(t)$ may grow to infinity, or decrease to zero slower than any power of $t$; typical examples of slowly growing functions are $\ln t$, $\ln(\ln t)$, $\ln^2 t$; their respective reciprocals $1/\ln t$, etc., provide examples of slowly vanishing functions. 
The presence of a slowly varying function may change the normal diffusion behavior at $\alpha =1$ to  super-diffusion if $L(t)$ slowly increases and to sub-diffusion if it decreases. Only if $L(t)$ has a finite, non-zero limit normal diffusion is observed. This limit then is related to the value of the normal diffusion constant $D_1$ by $\lim_{t\to \infty}L(t) = D_1 M^2/(2 C(0))$.
   
Because any estimate of the exponent $\alpha$ -- be it on the basis of numerically or experimentally determined data -- will be always contaminated by errors, the identification of the slowly varying function presents a difficult problem; even more, its presence also may considerably impede the estimate of the exponent $\alpha$  as it leads to slightly curved graphs rather than to a straight line with a well-defined slope in a doubly logarithmic plot of the position variance versus time.

Using the Tauberian theorem \cite{F}, see also appendix \ref{T}, we find, as an immediate consequence of the asymptotic behavior of the diffusion coefficient at large times, the corresponding behavior of the Laplace transform $\hat{D}(z)$ at small, positive $z$ to be given by
\begin{equation}
\hat{D}(z) \sim z^{-\alpha} \Gamma(\alpha) L(1/z)\quad \text{for}\: z\to 0\:,
\label{Daz}
\end{equation} 
where $\Gamma(z)$ denotes the Gamma-function.
Putting this result into Eq.~(\ref{zDg}) we obtain for the Laplace transform of the friction coefficient 
\begin{equation}
\hat{\gamma}(z) = \frac{2 C(0)}{\Gamma(\alpha) M^2 z^{2-\alpha}} L^{-1}(1/z) -1\quad \text{for}\: z\to 0\\:.
\label{gza}
\end{equation}
For $\alpha \geq 2$, the constant term $-1$ dominates in the limit  $z\to 0$ and, hence, the time-dependent friction coefficient $\gamma(t)$ quickly converges to zero. We shall not further consider this extreme case of super-ballistic motion. For $0 < \alpha <2$  the constant term in Eq.~(\ref{gza}) can be neglected. We may then again employ the Tauberian theorem to obtain
\begin{equation}
\gamma(t) \sim \frac{2 C(0)}{M^2 \Gamma(\alpha) \Gamma(2-\alpha)} t^{1-\alpha} L^{-1}(t)\:.
\label{gta}
\end{equation}
Combining the asymptotic results for the time-dependent diffusion and friction coefficients we find for the limiting behavior of the product of these two functions a generalized asymptotic Einstein relation of the form
\begin{equation}
\lim_{t\to \infty} D(t) \gamma(t) = \frac{2 C(0)}{M^2 \Gamma(\alpha) \Gamma(2-\alpha)}\quad \text{for} \;\;0<\alpha<2\:.
\label{Dga}
\end{equation}
We want to stress that the slowly varying functions modifying the algebraic behavior of the diffusion and the friction coefficients compensate each other in the product entering the limit on the right hand side.  Therefore this relation may be used as the basis of a more precise estimate of the scaling exponent $\alpha$. 
The standard form of the Einstein relation is recovered for $\alpha =1$ yielding $D \gamma$ as the limit on the left hand side and $2 C(0)/M^2 = 2 k_B T^{\text{set}} /M$ on the right hand side. 

Finally we note that taking the Laplace transformation on both sides of Eq.~(\ref{mPFe}) one can obtain the following formal expression for the Laplace transform of the average momentum in the presence of an external force, $\langle \hat{P}(z) \rangle_e = \int_0^\infty dt e^{-z t} \langle P(t) \rangle_e$, 
\begin{equation}
\begin{split}
 \langle \hat{P}(z) \rangle_e & = \frac{\hat{F}_e(z)}{z+\hat{k}(z)}\\
&= \frac{\hat{C}(z) \hat{F}_e(z)}{C(0)}\:,
\end{split}
\label{PFKz}
\end{equation}  
where we denoted $\hat{F}_e(z)  = \int_0^\infty dt e^{-z t} F_e(t) $ and put $\langle P(0) \rangle_e =0$. Making use of $\langle \delta \hat{X}(z) \rangle_e =\int_0^\infty dt e^{-z t} X(t) \rangle = \langle \hat{P}(z) \rangle_e/z$ in combination with Eq.~(\ref{Dz}) we obtain upon an inverse Laplace transformation
\begin{equation}  
\langle \delta X(t) \rangle_e = \frac{2 M}{C(0)} \int_0^\infty ds D(s) F_e(t-s)\:,
\label{XDFt}
\end{equation}
and for a constant force $F_e(t) = F$
we find with $\int_0^t ds D(s) = \langle (\delta X (t))^2 \rangle$
\begin{equation}
 \langle \delta X(t) \rangle_e = \frac{2 M}{C(0)} \langle (\delta X (t))^2 \rangle F_e\:.
\label{XXF}
\end{equation}
This equality relating the average position dynamics in the presence of a weak constant external force to the equilibrium mean square  deviation of the position  is also known as generalized Einstein relation \cite{BG,BF}.   
It is valid for all times but, in contrast to the generalized asymptotic Einstein relation (\ref{Dga}),  does not explicitly contain the scaling exponent.

\section{Results}\label{Results}
\begin{figure*}[t]
\includegraphics[width=0.32\textwidth]{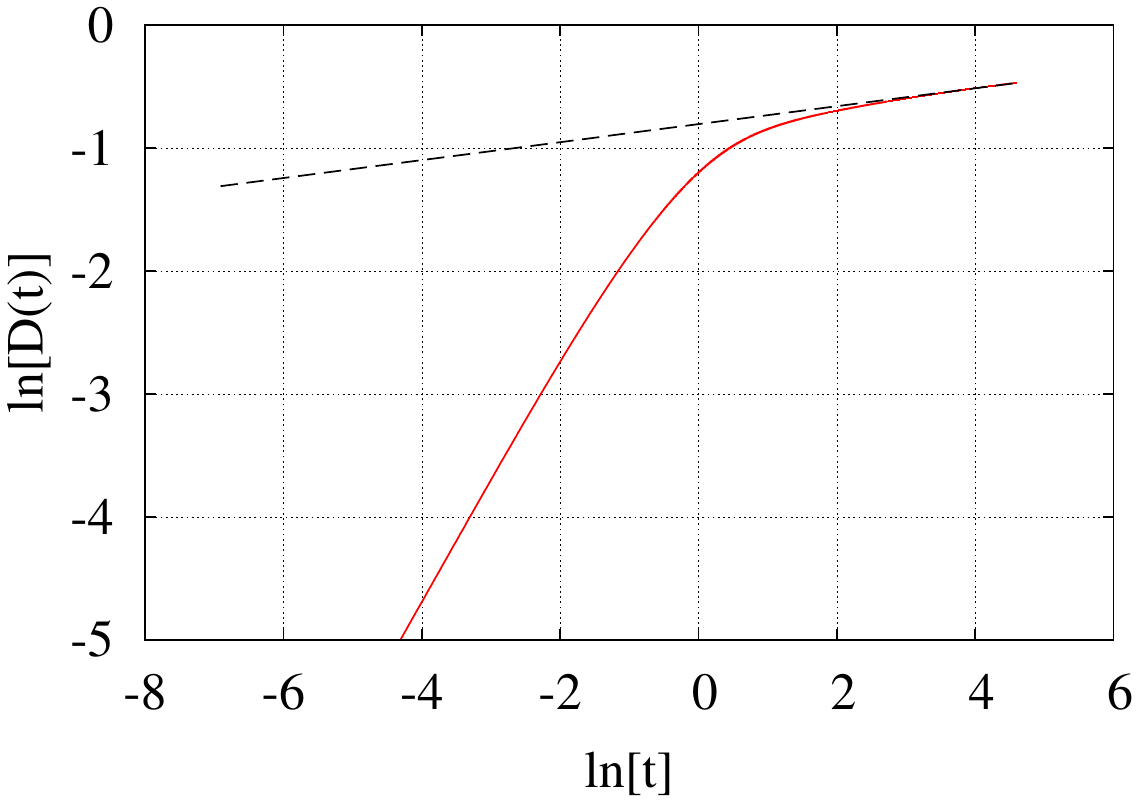}\nolinebreak
\includegraphics[width=0.32\textwidth]{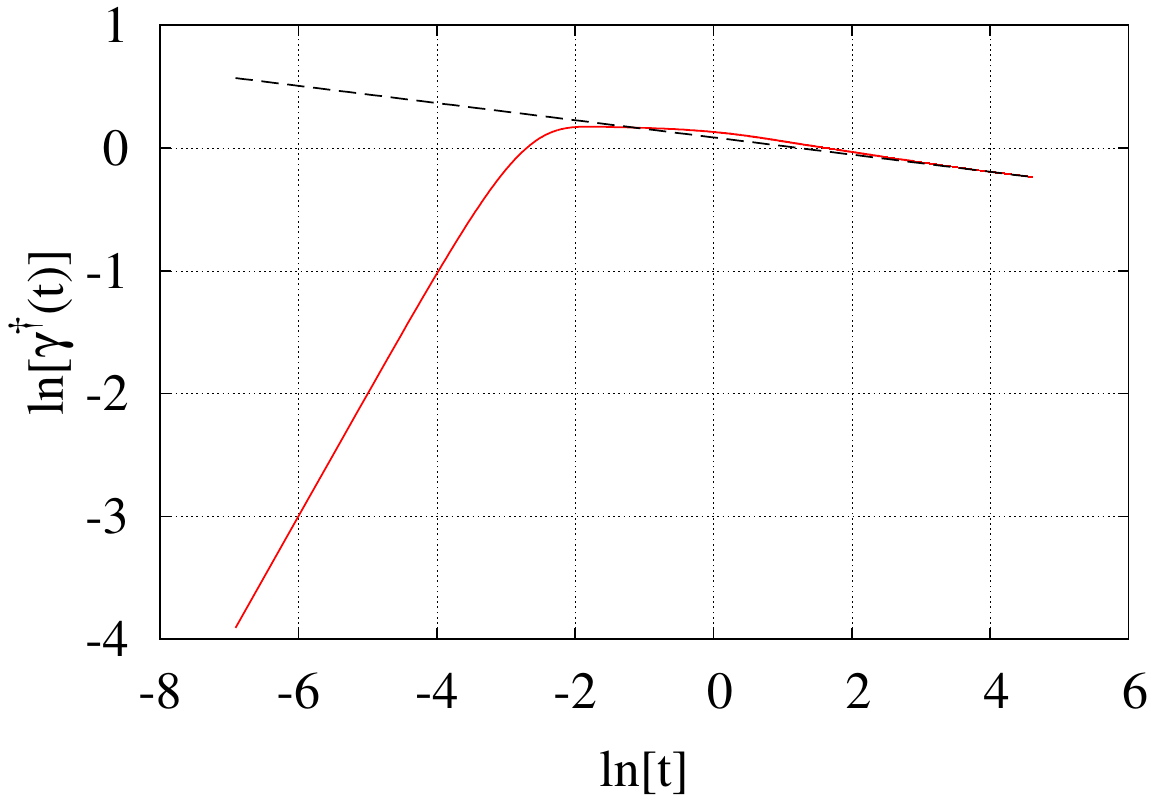}\nolinebreak
\includegraphics[width=0.32\textwidth]{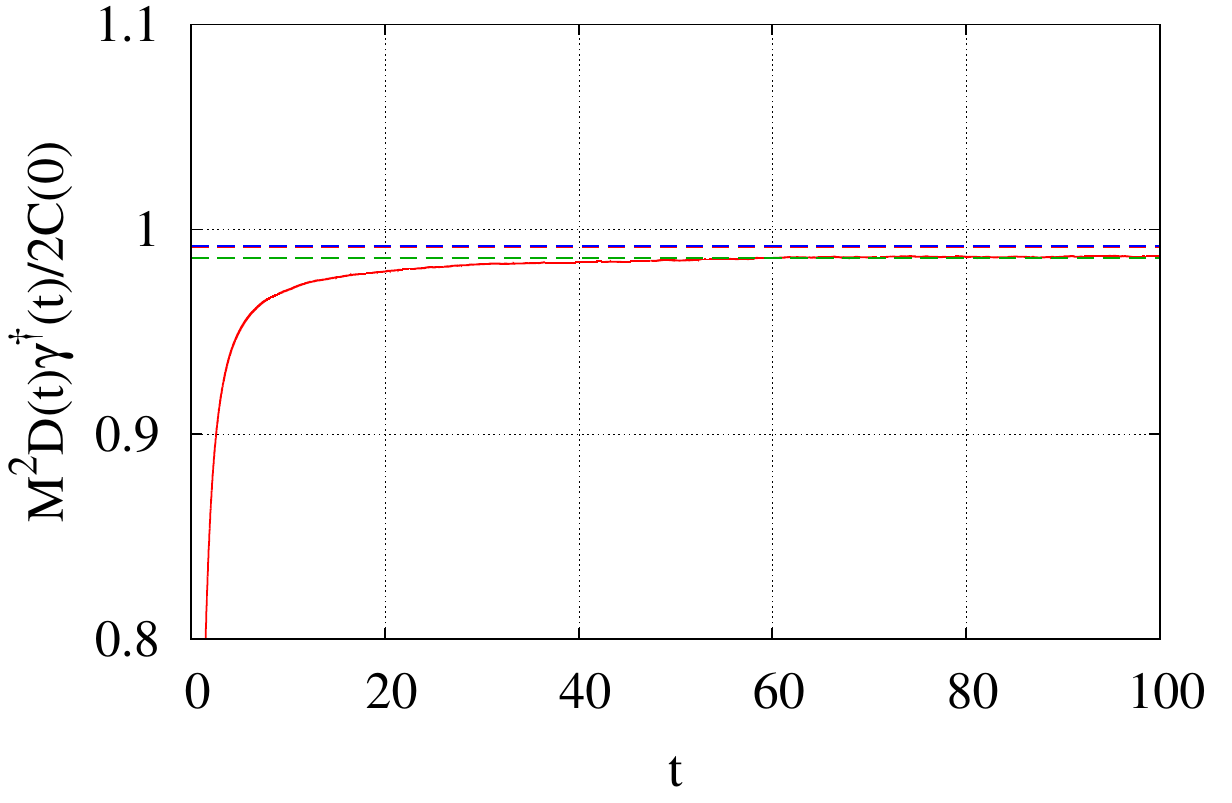}
\caption{Diffusion (left), friction (middle) and Einstein (right panel) coefficients  multiplied by $M^2/(2 C(0))$ for  self-diffusion ($M=m$, $\sigma_BB=\sigma_SS$) in a fluid consisting of $N=32\:000$ particles at density $n=0.3$ and at  temperature $T^{\text{set}}=0.5$ as functions of time. The red solid curves in the left and middle panel represent the diffusion and friction coefficients, respectively, determined from the molecular dynamics simulations. The scaling behavior of the diffusion coefficient is characterized at short times by $\alpha=2$ and corresponds to ballistic motion. The black broken straight lines in the left and middle panels are least square fits to the large time behavior of the diffusion and friction coefficients. From their respective slopes $s_D$ and $s_\gamma$ the estimates $\alpha_D=1+s_D$ and $\alpha_\gamma=1-s_\gamma$ of the scaling exponent can be obtained. For the results see Table~\ref{t1}. The red curve in the right panel again represents the scaled Einstein coefficient obtained from the molecular dynamics simulation. The green broken line displays the asymptotic value reached at large times. For comparison, the blue and the green broken lines, which almost fall on top of each other, indicate the values of the factor $1/(\Gamma(\alpha) \Gamma(2-\alpha))$ entering the generalized asymptotic Einstein relation (\ref{Dga}) for $\alpha=\alpha_\gamma$ and $\alpha=\alpha_D$, respectively. The results for the other two densities $n=0.4$ and $n=0.5$ as well as those for the higher temperature $T=1.0$ are qualitatively similar and therefore not shown.}
\label{scaling_ld}
\end{figure*}

We investigated the diffusion coefficient, the friction coefficient and the generalized asymptotic Einstein relation in the particular case of self-diffusion, i.e. for $\sigma_{BB} = \sigma_{SS}$ and $M=m$,  at different densities $n= 0.3,0.4, 0.5$ and 0.8 of the considered fluid. The corresponding temperatures were chosen as $T^{\text{set}} = 0.5,$ and 1.0 in the cases of the three lowest densities and $T^{\text{set}}=0.2$ in the high-density case.
In each case, the diffusion coefficient is determined from a numerical integration of the MACF according to the second line of Eq.~(\ref{Dt}).  The friction coefficient is obtained from the time-integral of the memory kernel. Both, the diffusion and the friction coefficients are displayed in doubly logarithmic plots in order to make a possible scaling behavior visible. In the case of the low densities, $n=0.3,\;0.4,\;0.5$, two scaling regimes can be distinguished for the diffusion coefficient, one for short times $t<1/e$ and the other one for large times $t>e$, and similarly for the friction coefficient reaching up to $t\approx e^{-3}$ for the short-time scaling regime and above $t\approx 1/e$ for the large-time scaling regime.
The maximal time $t\approx 100$ is sufficiently small that an influence from the finite sound velocity can be excluded for systems with $N=32 000$ particles.  
The left and the middle panel of Fig.~\ref{scaling_ld} display the diffusion and friction coefficient for $n=0.3$ and $T^{\text{set}}=0.5$.
The large time behavior of these coefficients can be described by (\ref{DaL}) and (\ref{gta}) with 
scaling exponents which almost agree with each other, see Table~\ref{t1}.
\begin{table}
\begin{tabular}[t]{|c|c||c|c|c|}
\hline
$T$& $n$&$\alpha^D$&$\alpha^\gamma$&$\alpha^{\text{ER}}$\\
\hline
\hline
0.5&0.3&1.073&1.070&1.092\\\hline
0.5&0.4&1.090&1.092&1.114\\\hline
0.5&0.5&1.10&1.098&1.118\\\hline
1.0&0.3&1.069&1.068&1.092\\\hline
1.0&0.4&1.087&1.085&1.096\\\hline
1.0&0.5&1.099&1.101&1.158\\\hline
0.2&0.8&1.013&1.019&1.032\\
\hline
\end{tabular}
\caption{Estimates of the scaling exponent $\alpha$ obtained from the behavior of $D(t)$, $\gamma(t)$ and the Einstein coefficient $D(t) \gamma(t)$ on the large-time range $[40,100]$ are collected for different temperatures and densities. In most cases, the difference between the values of the scaling exponents $\alpha^D$ and $\alpha^\gamma$ differ from each other only insignificantly. This indicates that, within the considered time interval $[40,100]$, both coefficients behave asymptotically according to eqs. (\ref{DaL}) and (\ref{gta}). In all cases, the exponent $\alpha^{\text{ER}}$ following from the generalized asymptotic Einstein relation is larger than the other two estimates. This suggests the presence of slowly varying functions which influence $\alpha^D$ and $\alpha^\gamma$ in the same way but cancel in the Einstein coefficient. }
\label{t1}
\end{table}
\begin{figure*}[t]
\includegraphics[width=0.32\textwidth]{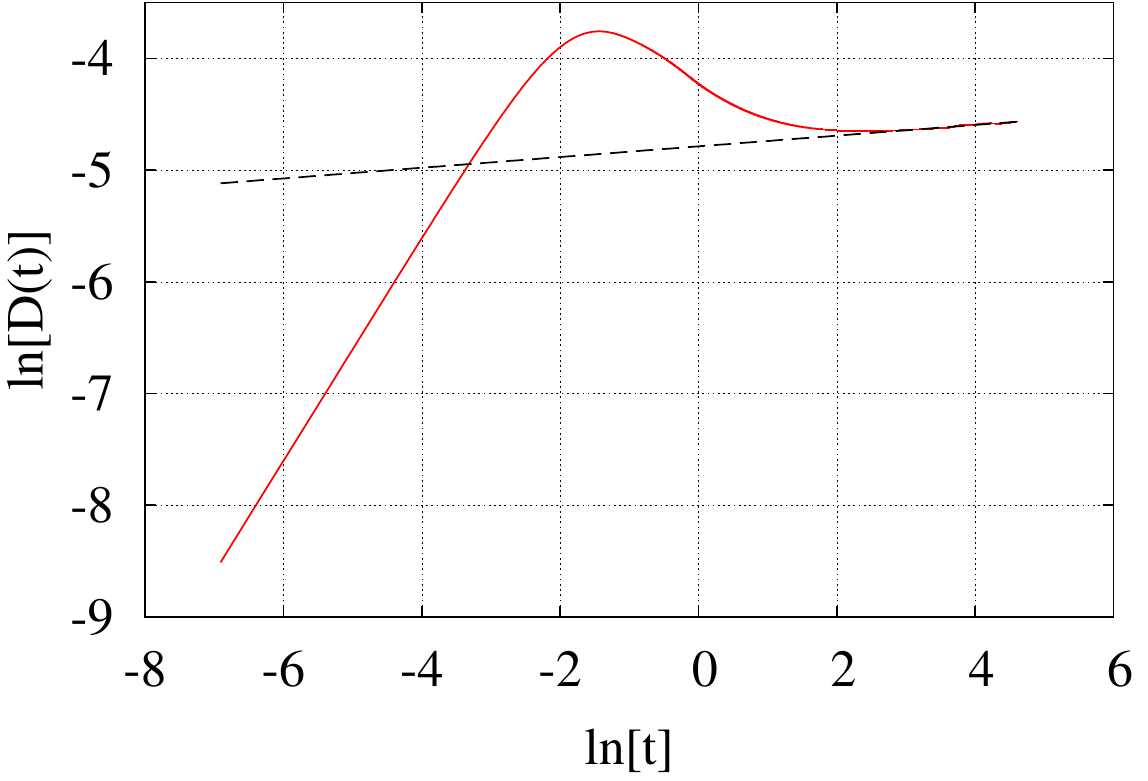}\nolinebreak
\includegraphics[width=0.32\textwidth]{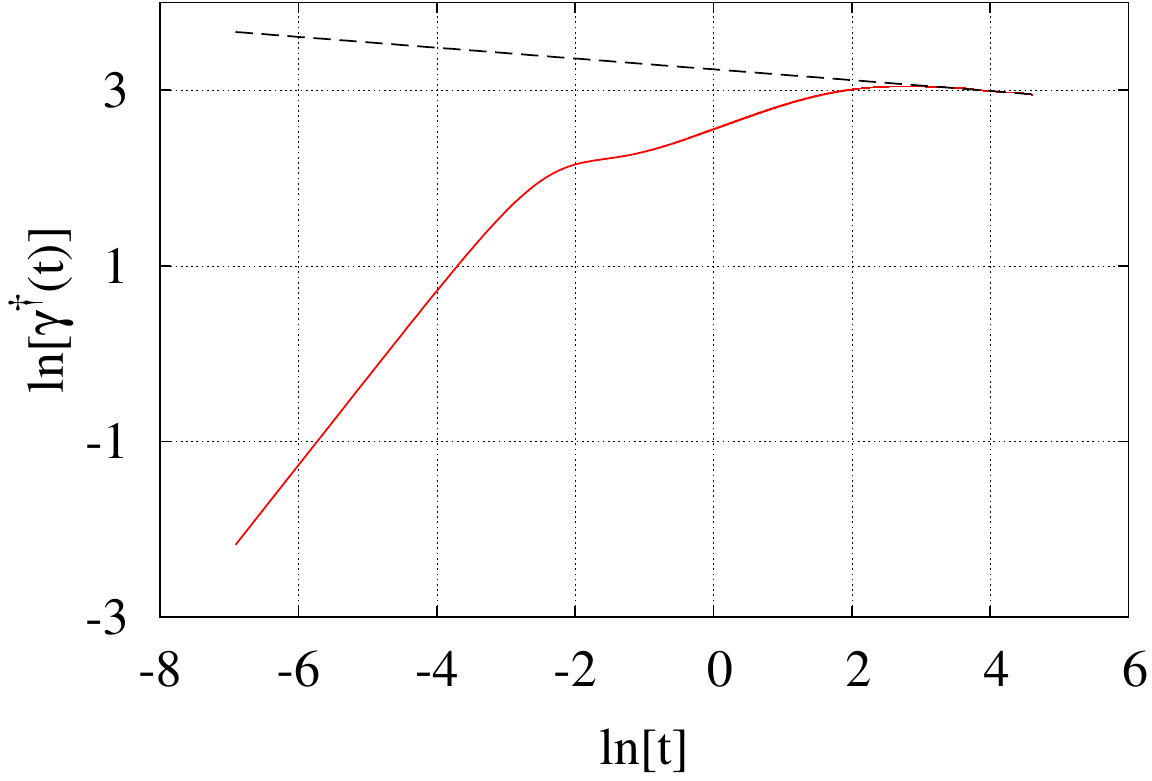}\nolinebreak
\includegraphics[width=0.32\textwidth]{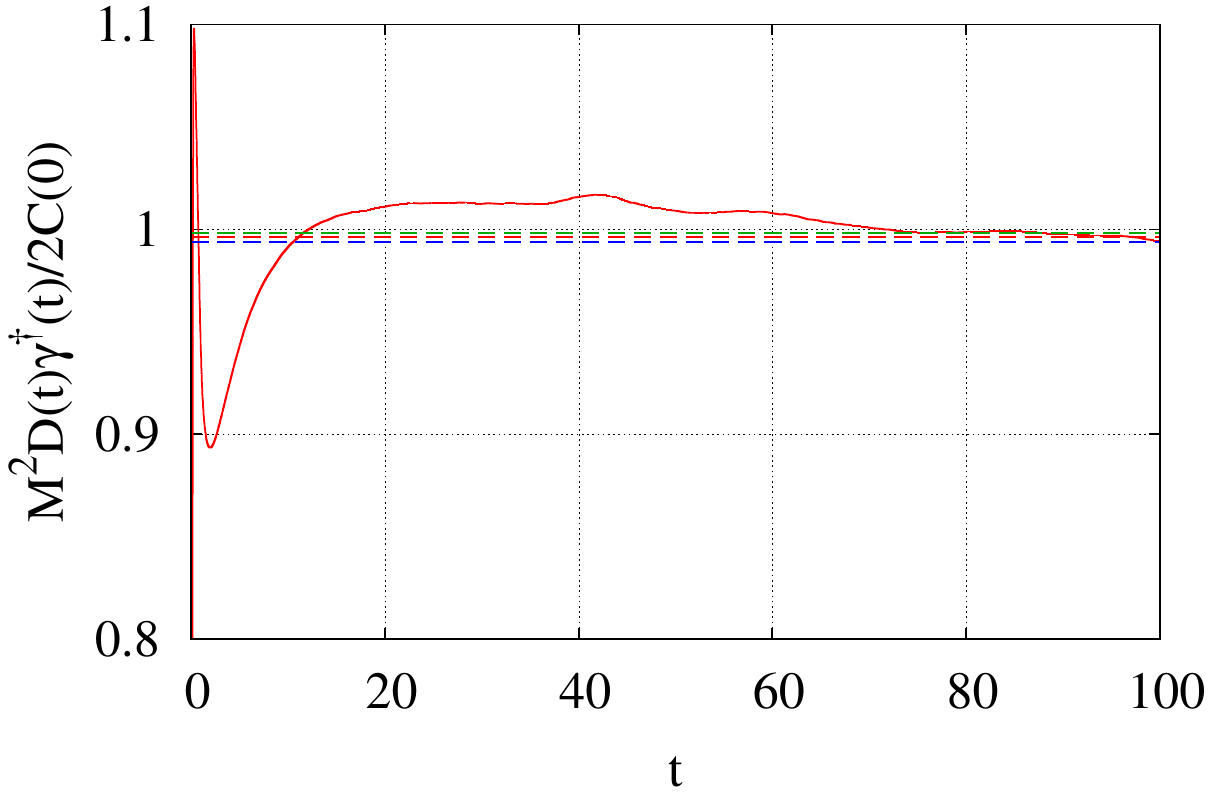}
\caption{Diffusion (left), friction (middle) and Einstein (right panel) coefficients for  self-diffusion ($M=m$) in a fluid with density $n=0.8$ and at the temperature $T^{\text{set}}=0.2$. Lines and color codes agree are the same as in Fig.~\ref{scaling_ld}. In the right most panel all three broken lines fall on top of each other.}
\label{scaling_hd}
\end{figure*}

In the regime of low densities $n=0.3,\;0.4,\;0.5$ the scaling exponent $\alpha$ is close to $1.1$ for both temperatures $T^{\text{set}}=0.5,\;1.0$. This  super-diffusive scaling behavior is different from the results of mode-coupling theory according to which the momentum autocorrelation decays as slowly as $\propto t^{-1}$ for a two-dimensional fluid \cite{AW} giving rise to a diffusion coefficient which grows $\propto \ln(t)$. The self-consistent mode-coupling  theory predicts the momentum auto-correlation function to decay $\propto t^{-1} (\ln(t))^{-1/2}$ \cite{K,WAG} and the diffusion coefficient to increase $\propto (\ln(t))^{1/2}$. Hence, according to both theories the scaling exponent would be $\alpha=1$ in disagreement with the present numerical findings.  The actually observed algebraic law with $\alpha \approx 1.1$  most likely is still modified by a slowly varying function. The precise nature of this slowly varying function is difficult to extract from the existing numerical data but its presence is strongly suggested by the fact that the apparent scaling exponents of the diffusion and the friction coefficients agree quite well with  each other, but not as accurately with the value of $\alpha$ following from the generalized asymptotic Einstein relation.
A detailed understanding of the mechanism underlying this behavior is still missing.

When going to the larger density $n=0.8$ the scaling exponent becomes approximately normal  with $\alpha \approx 1$, see Fig.~\ref{scaling_hd}. The 
asymptotic large time regime presumably is not yet fully reached as can be inferred from the larger difference of $\alpha_D$ and $\alpha_\gamma$ and also from the relatively large fluctuations of the Einstein coefficient at for times larger than $20$.  
An extension of the considered time-range would be problematic because of the sound velocity, which at this relatively high density is larger than in the low density cases. A compensation of this effect by considering larger systems at the same density would requires to consider even larger numbers of particles which makes the simulation very time-consuming.

\begin{figure}\begin{center}
\includegraphics[width=8cm] {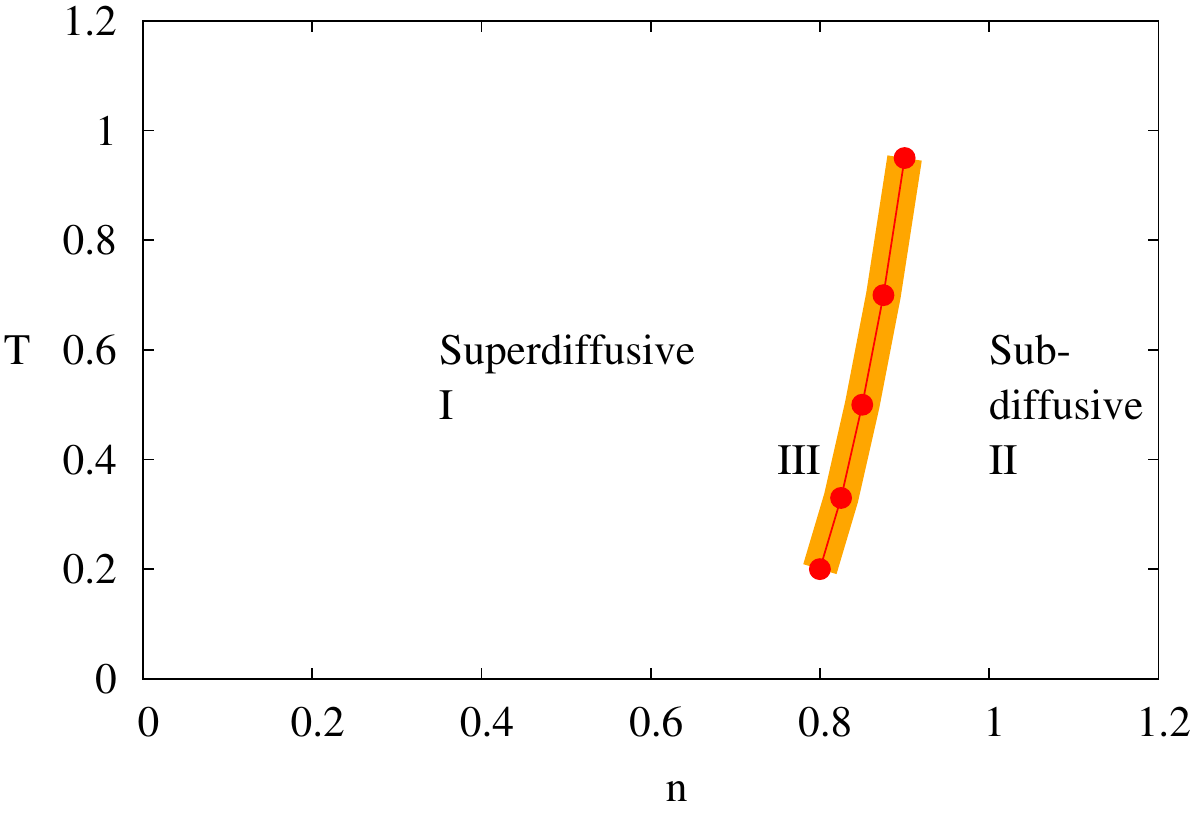}
\caption{The figure displays the density($n$)-temperature($T$) plane characterizing the self-diffusive motion in 2D soft-disk fluids. The yellow narrow region III separates the super-diffusive low-density fluid behavior I from a very slow, possibly sub-diffusive behavior II at large densities. Along this boundary, the self-diffusion is normal. Normal self-diffusive motion was confirmed by numerical simulations for density-temperature values corresponding to the 
indicated points within the narrow diffusive strip.} 
\label{Normal_diffusion_curve_n-T}
\end{center}\end{figure}

At even higher densities, for numerical reasons, it becomes  increasingly difficult to reach the large-time asymptotic regime. Preliminary results indicate a continuing tendency of the friction coefficient to increase with increasing time possibly giving rise to  sub-diffusive behavior. The Figure~\ref{Normal_diffusion_curve_n-T} roughly indicates the border region between super-and sub-diffusive motion in a density-pressure phase diagram.

\section{Conclusions}
Based on Mori's generalized Langevin equation for the momentum of a Brownian particle in combination with the assumption that its mean square displacement is of regular variation as a function of time, we derived a generalized asymptotic Einstein relation. It connects the time-dependent diffusion and the friction coefficients at large times for anomalously diffusing particles in an analogous way as the Einstein relation does for normal diffusion in a thermal environment. The only, but most relevant modification consists in a factor that multiplies the second moment of the momentum. This factor depends only on the scaling exponent of the considered anomalous diffusion process and becomes unity for normal diffusion. 
Here, the time-dependent diffusion and friction coefficients are defined as the derivative of the Brownian particle's mean square displacement with respect to time and the time-integral of the memory function entering the generalized Langevin equation, respectively.

The proof of the generalized asymptotic Einstein relation is based on the fact that,
under the given condition of a mean square displacement regularly varying at large times, the diffusion and friction coefficients are inversely proportional to each other, i.e. $D(t) \propto \gamma^{-1}(t) \propto t^{\alpha -1} L(t)$. 
Here we used that any function varying regularly in time may be represented as  $t^\alpha L(t)$ with $\alpha >0$ and $L(t)$ being slowly varying. 
A corresponding relation between the velocity autocorrelation function and the memory kernel has been known in the case of purely algebraic scaling, i.e. with a trivial, constant slowly varying function \cite{Corngold,Morgado,MCO}.

The use of the generalized asymptotic Einstein relation allows one to estimate the scaling exponent in a way that is not influenced by slowly varying functions.
We confirmed the inverse scaling behavior of the diffusion and the friction coefficients by means of molecular dynamics simulations of self-diffusion in a two dimensional liquid. The apparent deviation of the scaling exponents found from the diffusion and friction coefficients on the one hand and the generalized asymptotic Einstein relation 
can be explained by the influence of a slowly varying function.

We identified a large region in the temperature-density plane with super-diffusive self-diffusion and a cross-over region where the self-diffusion is normal. 
A super-diffusive  behavior was found by Isobe \cite{Isobe} for a hard-disk fluid at moderate densities. The scaling exponents found by Isobe are similar to the ones we found for a soft disk fluid.
The mechanism leading to the observed algebraic behavior is not known. In particular it is not explained by the existing mode-coupling theories \cite{AW,K,WAG}.

In the case of a two-dimensional Lorentz gas the existence of ``empty corridors'' is responsible for the occurrence of super-diffusion \cite{AHO}. Whether a similar mechanism may explain the more dynamic situation of a probe particle moving under the mutual influence of other particles presents an open question.   

The existence of a region with normal diffusion was reported by Liu, Goree and Vaulina \cite{LGV} for a system of particles mutually interacting via Yukawa potentials in two dimensions in a density-temperature region that is comparable with to the one where we observe normal self-diffusion. 
It is interesting to note that both the largest Lyapunov exponent and the Kolmogorov Sinai entropy have their maxima as functions of the density in the same region where one finds normal diffusion \cite{FP}. These quantities present measures of chaoticity of a system which is apparently largest for normal diffusion.

The derivation of the generalized asymptotic Einstein relation also holds for the motion of Brownian particles which may have mass and size that differ from those of the fluid particles. In contrast to other works, as in \cite{BMW}, in which  the Einstein relation is extended to non-equilibrium systems that still display normal diffusive behavior, our extension refers to equilibrium systems exhibiting anomalous diffusion. The generalized asymptotic Einstein relation must also be distinguished from another form of generalized Einstein relation that expresses the mean position response on a small constant force in terms of the particle's mean square displacement in thermal equilibrium \cite{BG,BF}.     

In the numerical confirmation of the generalized asymptotic Einstein relation we  restricted ourselves to self-diffusion, i.e. to the motion of a test-particle with the same mass and size as all other fluid particles.

\appendix
\section{Tauberian theorem}\label{T}
The Tauberian theorem relates the asymptotic behavior of a function $f(t)$ for large values of $t$ to the asymptotic behavior of its Laplace transform $\hat{f}(z) = \int_0^\infty e^{-z t} f(t)$ for small positive values of $z$. 
More precisely, if $f(t)$ is a monotone function for all $t >t_0\geq 0$ then
\begin{equation}
\hat{f}(z) \sim z^{-\rho} L(1/z) \quad \text{iff} \quad f(t) \sim \frac{1}{\Gamma(\rho)} t^{\rho -1} L(x)
\label{TT}
\end{equation}
where $0<\rho<\infty$ and $L(t)$ is a slowly varying function \cite{F}.  
\section*{Acknowledgement}
This research was supported by the Basic Science Research Program through the National Research Foundation of Korea funded by the Ministry of Science, ICT and Future Planning (Grant No. 2014-016000).


\begin{thebibliography}{99}
\bibitem{aniso} Even though the geometry of the configuration space is not strictly isotropic because of the slight difference of the lengths
$L_x$ and $L_y$ our simulations do not reflect any anisotropy.
\end{thebibliography}
\end{document}